\documentclass{aa}
\usepackage{graphicx}
\usepackage{natbib}
\usepackage{txfonts}

\newcommand{\methanol}{CH$_3$OH}
\newcommand{\htwoo}{H$_2$O}
\newcommand{\Lsun}{L$_\odot$}
\newcommand{\Msun}{M$_\odot$}
\newcommand{\cc}{cm$^{-3}$}

\begin{document}

\title{The Masers Towards IRAS~$20126+4104$}

\author{K. A. Edris\inst{1,2}, G. A. Fuller\inst{2},
  R. J. Cohen\inst{3} \and S. Etoka\inst{3}}

\institute{
Al-Azhar University, Naser City, Cairo, Egypt\\
\and
Physics Dept., UMIST, PO Box 88, Manchester, M60 1QD, UK\\
\and
University of Manchester, Jodrell Bank Observatory, Macclesfield, Cheshire SK11 9DL, UK}
 \date{Received 20 August 2004; accepted 22 December 2004}
\authorrunning{Edris et al.}
\offprints{G.Fuller@umist.ac.uk}

\abstract{ We present MERLIN observations of OH, water and methanol
  masers towards the young high mass stellar object IRAS~$20126+4104$.
  Emission from the 1665-MHz OH, 22-GHz H$_2$O and 6.7-GHz CH$_3$OH
  masers is detected and all originates very close to the central
  source. The OH and methanol masers appear to trace part of the
  circumstellar disk around the central source.  The positions and
  velocities of the OH and \methanol\ masers are consistent with
  Keplerian rotation around a central mass of ${\sim}$ 5~\Msun.
  The water masers are offset from the OH and \methanol\ masers and
  have significantly changed since they were last observed, but still
  appear to be associated to the outflow from the source.  All the OH
  masers components are circularly polarised, in some cases reaching
  100 percent while some OH components also have linear polarisation.
  We identify one Zeeman pair of OH masers and the splitting of this
  pair indicates a magnetic field of strength $\sim11$ mG within
  $\sim0.5''$ (850 AU) of the central source. The OH and
  \methanol\ maser emission suggest that the disk material is
  dense, $n>10^6$ cm$^{-3}$, and warm, $T>125$K and the high
  abundance of \methanol\ required by the maser emission is
  consistent with the evaporation of the mantles on dust grains in
  the disk as a result of heating or shocking of the disk
  material.
   }

\maketitle

\section{Introduction}

The extreme brightness and compact size of masers make them
valuable probes of regions of high mass star formation.  However
it is not clear how, or indeed if, the different types of maser,
\htwoo\ , OH and \methanol, are related to each other towards
typical high mass protostars.  Studies of \htwoo\ and \methanol\
masers towards a number of sources have failed to find a coherent
view of their relationship (e.g. Beuther et al 2002), however
statistical studies of OH and \methanol\ masers suggest that their
associations can tell us about the evolutionary sequence of star
formation (e.g.  Caswell 1997; Szymczak, \& Gerard 2004).

While the presence of particular types of maser towards a source may
trace the evolutionary stage of the source, many sources show emission
from multiple types of masers. Since the different masers can require
different excitation conditions, these multiple species can be used as
high resolution probes of
different components of the circumstellar environment.
Theoretical models suggest that the spatial coincidence
of different maser transitions or maser species can be used to infer
the properties of the emitting material. For example, Cragg et al.
(2002) found that gas phase molecular abundance is the key determinant
of observable maser activity for both OH and \methanol\ molecules.
Detailed modelling of high resolution observations towards particular
sources can also provide value insights on the structure of the
circumstellar material. For example, models have shown that the
complexity of observed magnetic field structure in W75N can be
explained by the maser emission originating from different depths
within the protostellar disk (Gray et al. 2003).

With these possibilities in mind, we present here the results of a
study of the \htwoo\ , OH and \methanol\ masers associated with
the well studied high mass young source IRAS~$20126+4104$. This
source is located in the Cygnus X region at an estimated distance
of 1.7 kpc (Wilking et. al. 1989).  It has a luminosity of $10^4$
\Lsun\ and is perhaps the best studied example of massive
protostar associated with a Keplerian disk and a jet/outflow
system. The molecular outflow has been mapped by Cesaroni et al.
(1997; C97) and observations at centimetre wavelengths have shown
that the outflow is fed by a jet (Hofner et al. 1999). On
the other hand, CH$_3$CN (5-4) observations (C97) have revealed a
molecular disk almost perpendicular to the jet axis and rotating
around the embedded young stellar object (YSO) at the origin of
the outflow/jet. Subsequent observations in the CH$_3$CN (12-11)
(Cesaroni et al. 1999; C99) and NH3 (1,1) lines (Zhang et al. 1998)
have found evidence for Keplerian rotation, implying a central disk
plus stellar mass of 24 \Msun.

The source is associated with OH, \htwoo\ and \methanol\ masers.
Two features of OH masers were detected in the 1665-MHz line by
Cohen et al. (1988) and more recent unpublished VLA data (Cohen,
priv. comm.). Observations of the water masers using the VLA with
angular resolution of $0.1''$ identified three emission regions
(Tofani et al. 1995). Moscadelli et al. (2000; hereafter MCR)
resolved two of these into 26 unresolved spots using the VLBA. The
velocity and spatial structure of these spots were well fitted by
a model with the spots arising at the interface between a jet and
the surrounding molecular gas. Maser emission from the 6.7-GHz
line of \methanol\ has recently been observed by Minier et al.
(2001). Using the EVN, they found two maser clumps separated by
100 AU and $0.8''$ to the north of the central source, a region
relatively remote from the central source and with no other known
indication of activity related to star formation.

To determine how these three types of masers are related to the
circumstellar disk and/or the jet and investigate the connection
between them, we have observed IRAS~$20126+4104$ at high angular
resolution using MERLIN.  The details of the observations and
reduction are given in Sec. 2 and the results presented in Sec. 3. In
Sec. 4 we discuss the interpretation while conclusions are drawn in
Sec. 5.

\begin{table*}
\begin{tabular}{|c|c|c|c|} \hline

  {Type of maser}  & {OH} & {H$_2$O} &{CH$_3$OH} \\ \hline
  Date of observation & 20  & 28 \& 29  & 16 \& 17  \\
                      & Jan 2002 & Mar 2002 & May 2002 \\
  Antenna Used & De CA DA  & CA MK2 DA  & CA MK2 \\
               & KN MK2 TA & KN TA       &       \\
  Field centre (2000) & $\alpha  20^{h} 14^{m} 26.04^{s}$ & $\alpha  20^{h} 14^{m} 26.04^{s}$ & $\alpha  20^{h} 14^{m} 26.04^{s}$ \\
                      &$\delta  41^{\circ} 13' 32.5''$ & $\delta  41^{\circ} 13' 32.5''$ & $\delta  41^{\circ} 13' 32.5''$ \\
  Rest frequency (MHz) & 1665.402 1667.359 & 22235.0798  & 6668.518  \\
  No. of frequency channels & 1024 & 256 & 512 \\
  Total band width (MHz) & 1 & 4 & 2 \\ 
  bandpass calibrator & 3C84 & 4C39.25 & 3C84 \\
  Polarisation angle calibrator & 3C286 & No & No \\
  Phase calibrator & 2005+403 & 2005+403 & 2005+403 \\ \hline
\end{tabular}
\caption{Observing and calibration parameters for the MERLIN
spectral-line observations of IRAS~$20126+4104$}
\end{table*}

\section{Observations and data reduction}

Table 1 gives the parameters for the MERLIN observations. All
measurements used the same phase calibrator source 2005+403 to
retrieve the absolute position of the maser spots and therefore
compare their locations from one line to another with high accuracy. A
bandpass calibrator was observed to calibrate the variation of
instrumental gain and phase across the spectral bandpass. For OH,
observations of 3C286 were also made during the observing run, with
the same correlator configuration and bandwidth, to calibrate the
polarisation characteristics. The data were reduced in Jodrell Bank
observatory using the MERLIN d-programs and the AIPS software package.

\subsection{OH masers}

IRAS~$20126+4104$ was observed in the 1665- and 1667-MHz OH maser
transitions in January 2002 using six telescopes of the MERLIN
network. The frequencies were alternated during the
observations, cycling between the two OH line frequencies, to
provide data on both transitions spread over the whole observing
track. The velocity resolution was 0.42 km s$^{-1}$ for a total of
1 MHz spectrum bandwidth corresponding to 180 km s$^{-1}$ velocity
range. The left- and right-hand circular (LHC and RHC)
polarisation data for each baseline were simultaneously correlated
in order to obtain all Stokes parameters. Using d-programs (see
Diamond et al. 2003), the data were edited and corrected for
gain-elevation effects. The flux density of the amplitude
calibrator 3C84, was determined by comparing the visibility
amplitudes on the shortest baselines with those of 3C286. Using
flux densities of 13.6 Jy at 1665 MHz and 1667 MHz for 3C286
(Baars et al. 1977), the flux density of 3C84 at the time of the
observation was determined to be $23.2\pm0.6$ Jy.

In AIPS the data were calibrated for all remaining instrumental
and atmospheric effects. Starting from a point source model, the
phase calibrator source was mapped, with a total of three rounds
of phase self-calibration and the resulting corrections applied to
the source data.  The polarisation leakage for each antenna was
determined using 3C84 and the polarisation position angle
correction was performed using 3C286. The AIPS task IMAGR was used
to map the whole data set in Stokes I, Q, U and V in order to
retrieve the polarisation information. The rms noise, after
CLEANing, was typically 14 mJy/beam and the FWHM of the restoring
beam is 174 $\times$ 137 mas at a position angle of $-41^\circ$.

The positions of the maser components were determined by fitting
two-dimensional Gaussian components to the brightest peaks in each
channel map. Components were considered as spectral features if
they occurred in three or more consecutive channels. Using flux
weighted means over those channels of each group, the positions
and velocities of spectrum features were obtained. The
uncertainties in relative positions are typically 10 mas.

\subsection{Water masers}

The \htwoo\ maser line at 22GHz was observed using 4C39.25 as
bandpass and flux calibrator.  The flux density of 4C39.25 at this
frequency was taken to be 7.8 Jy (Terasranta priv.  comm.).  The
phase calibrator was mapped with a total of two rounds of phase
self-calibration and the resulting corrections applied to the
IRAS~$20126+4104$ data. The spectral bandwidth was 4 MHz
corresponding to 54 km s$^{-1}$ velocity range with channel
separation of 0.25 km s$^{-1}$. Maps of all the spectral channels
were generated and de-convolved using the AIPS task IMAGR. The
restoring data beam had a FWHM of 40 $\times$ 8 mas at a position
angle of $-37^\circ$. The rms noise was typically 11 mJy/beam but
up to 40 mJy/beam in the spectral channels with the brightest
emission.

\subsection{\methanol\ masers}

The 6.7-GHz methanol line was observed with just the two antennas
in the MERLIN array equipped with the appropriate receivers at the
time of the observations.  The correlator was configured to give a
velocity resolution of 0.21 km s$^{-1}$ for a total of 2 MHz
spectrum bandwidth corresponding to 90 km s$^{-1}$ velocity range.
3C84 was used as the bandpass and amplitude calibrator.  Its
amplitude at the time of the observations was estimated to be
16.5$\pm$1.6~Jy (Richards, private communication). The phase
calibrator source was mapped with one round of phase
self-calibration followed by an amplitude and phase self calibration.
Again IMAGR was used to make the images and CLEAN them. The resulting
FWHM of the beam is 26 $\times$ 24 mas at a position angle of
$39^\circ$.

\subsection{Positional Uncertainties}

The accuracy of the absolute masers position measured in the paper
is limited by four factors : (1) the position accuracy of the
phase calibrator, (2) the accuracy of the telescope positions, (3)
the relative position error depending on the beamsize and
signal-to-noise ratio and (4) finally the atmospheric variability
that plays an important role especially at 22~GHz depending on the
angular separation between the calibrator (2005+403) and the
target. The first two factors are frequency independent and were
estimated to be 5 mas (given by the MERLIN calibrator catalogue)
and 10 mas respectively (Diamond et al. 2003). The other two
factors are frequency dependent. The relative position error,
given approximately by the beamsize/signal-to-noise ratio, leads
to uncertainties in the position of 17, 4 and 2.5 mas at OH,
\htwoo\ and \methanol\ maser lines respectively. Factor (4) is
inferred from the quality of the phase of the calibrator. With a
separation of $2^\circ$ between the phase calibrator and the
target this factor adds an uncertainties in the absolute position
of 5 to 20 mas at the 1665-MHz OH maser line for the typical and
worst phase rate respectively. For the \htwoo\ and 6.7-GHz
\methanol\ maser lines, it adds errors of 10 and 2 mas
respectively, taking into account the worst phase rate for each
observation. All these uncertainties combine quadratically to give
absolute position errors of 25, 15 and 12 mas in the 1665-MHz OH,
22-GHz \htwoo\ and 6.7-GHz \methanol\ component positions
respectively.


\begin{table*}
\begin{tabular}{|c|c|c|c|} \hline
  {Type of }  & {$\alpha$ (J2000)} & {$\delta$ (J2000)}&  {Velocity}      \\
  {maser   }  &                    &                   &  {(km s$^{-1}$)} \\ \hline
  {OH (RHC)}       & $ 20^{h} 14^{m} 26^{s}.060\pm0^{s}.002$ & $
41^{\circ} 13' 32''.63\pm0''.02$  & -12.1 \\
\hline
  {H$_2$O}         & $ 20^{h} 14^{m} 26^{s}.022\pm0^{s}.001$ & $ 41^{\circ} 13' 32''.60\pm0''.01$  & -7.1  \\
                   \hline
  {CH$_3$OH}       & $ 20^{h} 14^{m} 26^{s}.051\pm0^{s}.001$ & $ 41^{\circ} 13' 32''.70\pm0''.01$  & -6.1  \\
                    \hline

\end{tabular}
\caption{The absolute positions and velocities of the brightest
maser spots for the three maser types.}
\end{table*}

\begin{figure*}
  \centering
  \includegraphics[angle=0,width=10cm]{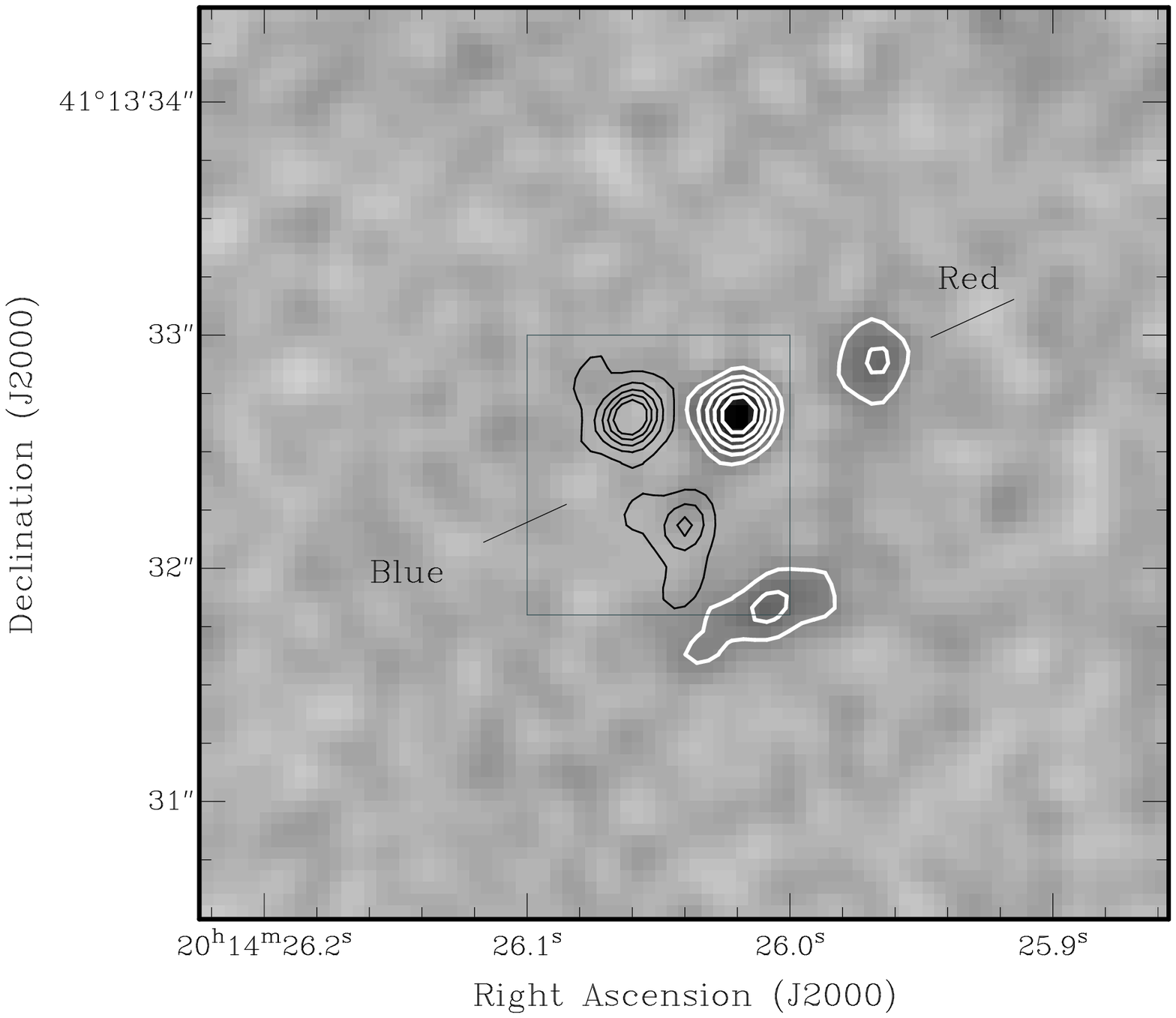}
  \includegraphics[angle=-90,width=18cm]{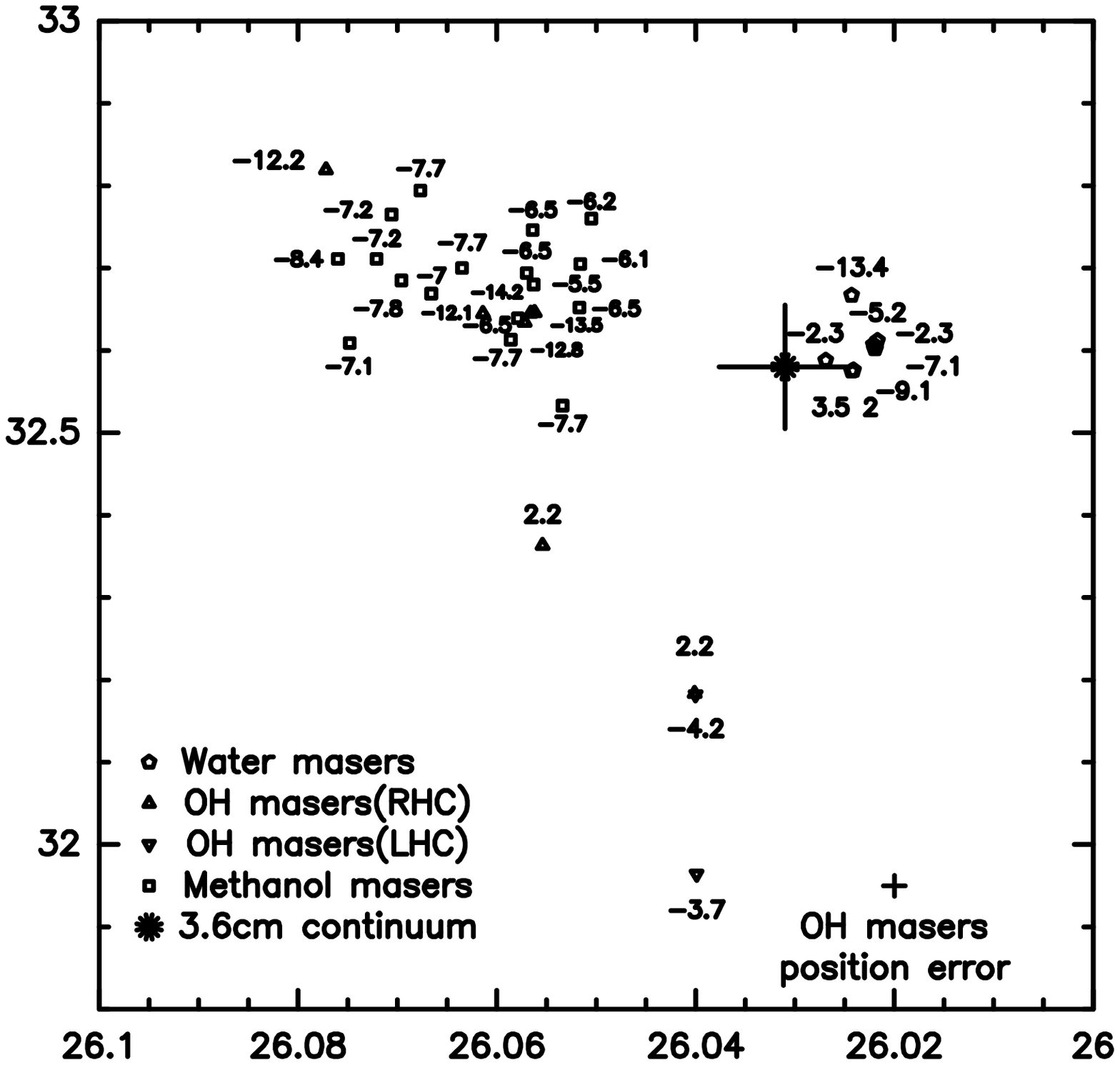}
\caption{Top panel: OH maser emission from IRAS~20126+4104 (thin
  contours) superimposed on an image of the 3.6-cm continuum
  emission (also bold contours) from Hofner et al. (2004), aligned in
  J2000 coordinates.
  The OH masers peak flux is 13.3 Jy/beam
  and the contour levels of the map are 1, 3, 5, 7 and 9 Jy/beam.
  The straight lines refer to the jet/outflow directions (C99), while the box
  refer to the location of the bottom panel.
  Bottom panel: positions and velocities of the OH, H$_2$O and methanol maser
  components and the position of the 3.6cm peak with accuracy bar (Hofner
  priv. comm.). The cross bars in the bottom right corner show OH masers
  position uncertainty. Note the close association of the three masers
  species and particularly between the OH and methanol masers.
}\label{fig:oh_radio_position}
\end{figure*}

\section{Results}

The 1665-MHz OH line, water and 6.7-GHz methanol masers were
detected with MERLIN. Table 2 gives the absolute positions and
velocities of the brightest maser spot for each maser type.
Radial velocities, here and elsewhere, are given relative to the
Local Standard of Rest (LSR). All three types of maser are
located close to the central source. The details of the emission
in each maser type are discussed below. The 1667-MHz OH maser line
was not detected down to a noise level of 20 mJy/beam.


\begin{table*}
\begin{tabular}{|c|c|c|c|c|c|c|c|} \hline

{No.} & {Vel.} &   {Flux} & {RA}     &  Error  & {DEC} & Error
& Notes  \\
   &    km s$^{-1}$   &   Jy/b     & s        &    s    &   $''$      &   $''$  &   \\ \hline
R comp. & & & & & & &  \\
 1 &   2.17  &   1.51$\pm$0.02 & 26.0401  &  0.0001 &  32.183 &
 0.002 & Z \\
 2 &   2.19  &   0.17$\pm$0.02 & 26.0554  &  0.0015 &  32.363 &
 0.032 & \\
 3 &  -12.10 &   2.70$\pm$0.03 & 26.0614  &  0.0011 &  32.645 &
 0.012 & \\
 4 &  -12.16 &   0.34$\pm$0.03 & 26.0772  &  0.0005 &  32.819 &
 0.006 & \\
 5 &  -12.80 &   0.32$\pm$0.02 & 26.0572  &  0.0007 &  32.634 &
 0.008 & \\
 6 &  -13.54 &   0.43$\pm$0.02 & 26.0562  &  0.0002 &  32.646 &
 0.003 & \\
 7 &  -14.21 &   0.32$\pm$0.02 & 26.0566  &  0.0003 &  32.645 &
 0.004 & \\
L comp. &  & & & & & & \\
 1 &  -3.75  &   0.96$\pm$0.02 & 26.0399  &  0.0001 &  31.964 &
 0.001 &  \\
 2 &  -4.16  &   0.62$\pm$0.02 & 26.0400  &  0.0006 &  32.182 &
 0.010 & Z \\ \hline

\end{tabular}
\caption{The parameters of the left and right hand circular
 polarisation components of 1665-MHz OH masers detected towards
 IRAS~$20126+4104$. The label Z marks a Zeeman pair. The leading terms
 of the positions are $\alpha$(J2000)= 20$^h$ 14$^m$ and $\delta$(J2000)= 41$^{\circ}$ $13'$.}
\end{table*}

\begin{table*}
\begin{tabular}{|c|c|c|c|c|c|c|c|c|c|c|c|c|} \hline

{No.}& {Vel.} &    {I} &  {Q}     &   {U}    &     {V}  &  {P} &{$\chi$}&{m$_L$}&{m$_C$} & {m$_T$}\\
   &km s$^{-1}$&   Jy/b&  Jy/b    &  Jy/b    &    Jy/b  & Jy/b &$^{\circ}$&\%
   &  \%    &   \%    \\ \hline
 1 &   2.17  &   0.95  &   0.00   &   0.00   &     0.75 & 0.00 &  0.00  &  0.0
   &   78.8 &   78.8  \\
 2 &   2.19  &   0.11  &   0.00   &   0.00   &     0.06 & 0.00 &  0.00  &  0.0
   &   55.8 &   55.8  \\
 3 &  -3.75  &   0.56  &   0.20   &  -0.03   &    -0.38 & 0.20 & -5.07  & 36.4
   &  -68.1 &   77.2 \\
 4 &  -4.16  &   0.36  &   0.00   &   0.00   &    -0.24 & 0.00 &  0.00  &  0.0
   &  -65.8 &   65.8  \\
 5 & -12.10  &   1.89  &   0.07   &  -0.03   &     1.34 & 0.08 &-11.11  &  4.1
   &   70.8 &   70.9 \\
 6 & -12.16  &   0.21  &   0.02   &   0.00   &     0.17 & 0.02 &  0.00  & 11.5
   &   78.4 &   79.3  \\
 7 & -12.80  &   0.14  &$<$0.02   &   0.00   &     0.16 & 0.00 &  0.00  &  0.0
   &  100.0 &  100.0  \\
 8 & -13.54  &   0.24  &$<$0.02   &   0.00   &     0.21 & 0.00 &  0.00  &  0.0
   &   86.6 &   86.6  \\
 9 & -14.21  &   0.16  &   0.00   &   0.00   &     0.14 & 0.00 &  0.00  &  0.0
   &   85.7 &   85.7  \\ \hline

\end{tabular}

 \caption{The Stokes and polarisation parameters of the 1665-MHz OH
masers components detected towards IRAS~$20126+4104$. }
\end{table*}

\subsection{Hydroxyl masers}

A total of 9 (7 RHC and 2 LHC) 1665-MHz OH maser spots were
detected towards IRAS~$20126+4104$.  Table 3 presents the
parameters of the OH maser components detected, namely the peak
intensities, velocities and positions for each hand of circular
polarisation. The label Z marks a left-hand and right-hand
polarised pair of components which originate at the same location
and are identified as a Zeeman pair with a splitting of 6.3 km
s$^{-1}$. Figure~\ref{fig:oh_radio_position} shows the
distribution of the OH maser spots.  They are distributed in two
clusters on opposite sides of the axis inferred for the radio
continuum jet (C99, Hofner et al. 1999) and about $0.3''$
south-east from the 3.6cm continuum centre. The OH maser spots are
spread over a region of $\sim1.2''$ corresponding to 2000 AU (or
$\sim$ 0.01 pc) at a distance 1.7 kpc. The distribution of OH
masers is approximately symmetrical about a line of NE-SW
direction at position angle 29$^\circ$.
Figure~\ref{fig:oh_radio_position} also shows the spatial
distribution of the velocities of the maser components. For the OH
masers the velocities are mainly negative to the north and
positive to the south.

Table 4 present the Stokes parameters I, Q, U and V, the
polarisation position angle ($\chi$) (angles are measured from N
towards E), the linearly polarised flux P, the percentage of
linear polarisation $m_L$, the percentage of circular polarisation
$m_C$ and the total percentage of polarisation $m_T$ of each
feature.  The OH masers have a total polarisation around
80\% ranging from 55\% to 100\%. The Stokes intensities are shown
as zero in this table if their flux is below the noise level. All
1665-MHz features are circularly polarised and three features (3,
5 and 6) are elliptically polarised. Feature 3 is the most
elliptically polarised (36.4\%) and feature 7 is 100 \% circularly
polarised. Polarisation position angles can only be measured for
features 3 and 5.

\subsection{Water masers}

\begin{figure}
  \centering
  \includegraphics[angle=0,width=6cm]{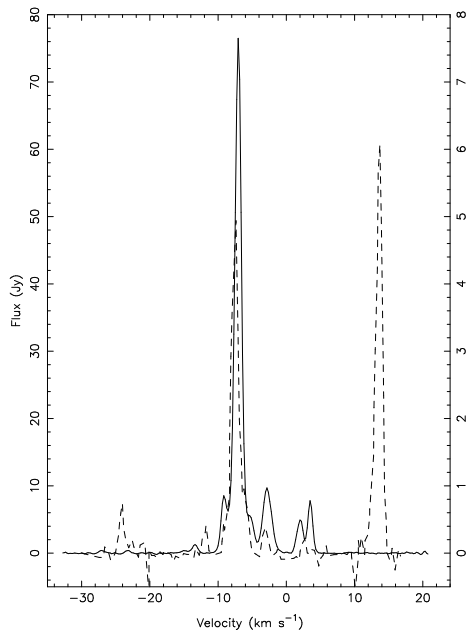}
\caption{H$_2$O masers spectra towards IRAS~$20126+4104$ from
MERLIN observations in March 2002 (solid line) compared with that
of Medicina single dish observations in Nov. 1997 (dashed line)
from MCR. MERLIN detected no emission between velocities 10 to 16
km s$^{-1}$ while the component centred at -7.1 km s$^{-1}$ became
very bright. The axis on the left side shows MERLIN intensity and
the axis on the right side shows the Medicina
intensity}\label{fig:22spec}
\end{figure}

\begin{table}
\begin{tabular}{|c|c|c|c|c|}
  \hline

  {NO.} & V$_{LSR}$ & Flux & $\alpha$ & $\delta$      \\

        &km s$^{-1}$& Jy/beam        & sec      & $''$            \\\hline
     1  &   3.51    &  7.79$\pm$0.01 & 26.0243  & 32.575       \\
     2  &   2.05    &  4.25$\pm$0.01 & 26.0241  & 32.576       \\
     3  &  -2.27    &  8.40$\pm$0.01 & 26.0269  & 32.588       \\
     4  &  -2.36    &  0.56$\pm$0.01 & 26.0221  & 32.607       \\
     5  &  -5.22    &  4.93$\pm$0.01 & 26.0217  & 32.612       \\
     6  &  -7.07    & 78.81$\pm$0.04 & 26.0219  & 32.602       \\
     7  &  -9.14    &  8.53$\pm$0.01 & 26.0220  & 32.602       \\
     8  & -13.39    &  1.20$\pm$0.01 & 26.0243  & 32.667       \\  \hline

\end{tabular}
\caption{The parameters of the eight maser components detected in H$_2$O masers
 towards IRAS~$20126+4104$. The position errors are $0^s.0003$ for $\alpha$
 and $0.''004$ for $\delta$. The leading terms  of the positions are
 $\alpha$(J2000)= 20$^h$ 14$^m$ and $\delta$(J2000)= 41$^{\circ}$ $13'$.}
\end{table}

\begin{figure}
  \centering
  \includegraphics[angle=270,width=9cm]{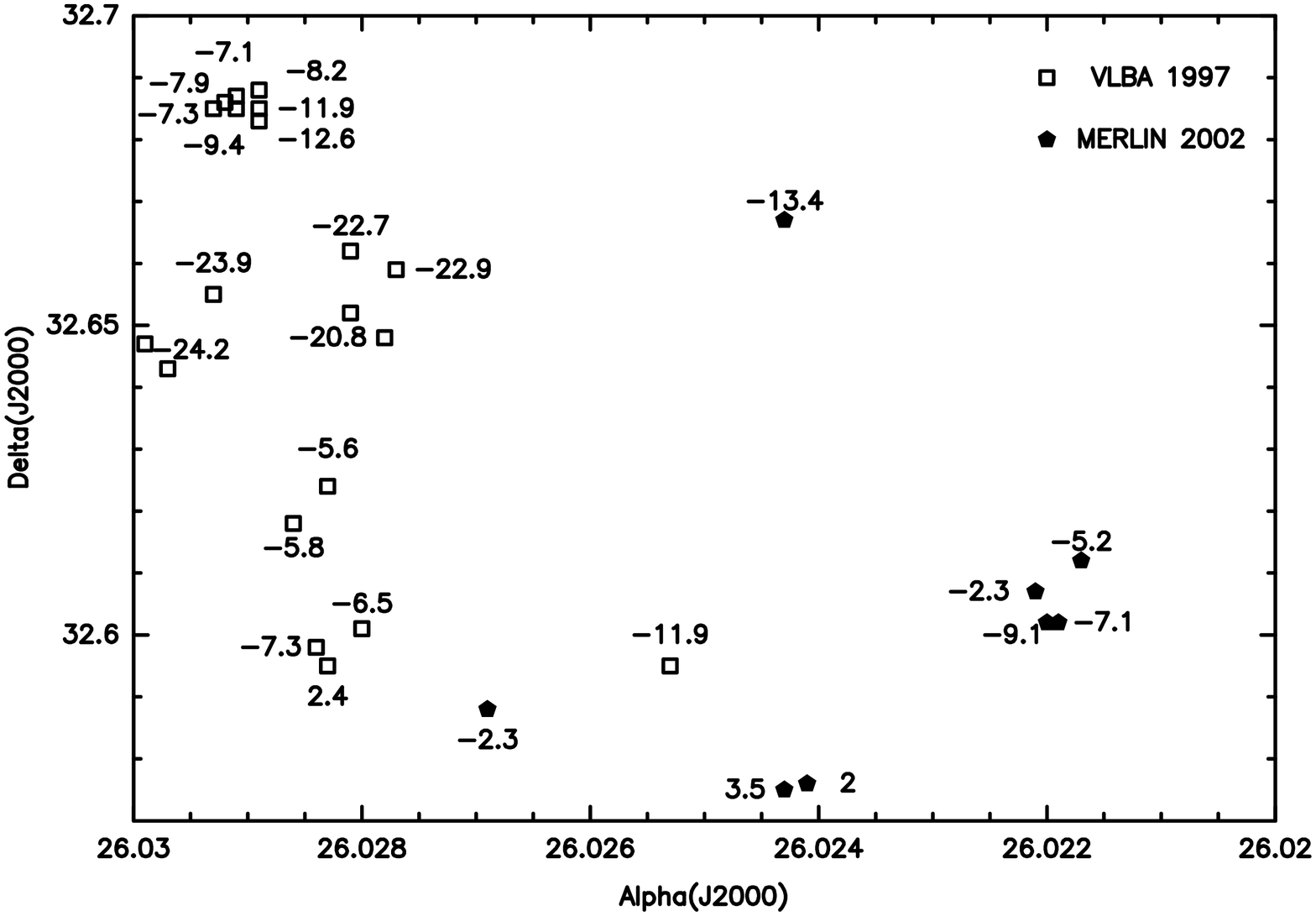}
\caption{Comparison between the water masers positions from our MERLIN
  and the VLBA observations towards IRAS~$20126+4104$. Between these
  observations the masers spots have either changed almost completely
  or perhaps have moved to the south west.  MCR maser spots located
  very far from the MERLIN spots are not shown.}\label{fig:h2o_comp}
\end{figure}

The water masers towards IRAS~$20126+4104$ were detected in a
single cluster of $\sim0.15''$ (255 AU at 1.7 kpc) in size close
to the central source and the OH (and methanol, \S~\ref{sec:meth})
masers. This cluster is close to the C2 group of spots identified
by Tofani et al. (1995). However the positions and fluxes have
varied strongly, the brightest emission being 20 times greater
than seen by Tofani et al.  Also, neither of the clusters C1 and
C3 identified by Tofani et al. was detected with MERLIN, implying
that they have decreased in brightness by factors of more than 10
and 100 respectively.

Figure~\ref{fig:22spec} shows that the spectrum of the water
masers detected with MERLIN is also considerably different from
that observed by MCR. The strongest component detected by MCR
centred at $V_{LSR}$ $\sim 12$ km s$^{-1}$ is not detected with
MERLIN nor are the highest velocity components centred between -30
km s$^{-1}$ and 20 km s$^{-1}$. On the other hand, MERLIN detected
components between 0 km s$^{-1}$ and +5 km s$^{-1}$ which were
previously not detected with the VLBA.  At the same time, the
component centred in the velocity range -5.0 km s$^{-1}$\ and 0 km
s$^{-1}$ has increased in strength and the component centred at
$\sim-7$ km s$^{-1}$ dominates both spectra, although as
Figure~\ref{fig:h2o_comp} shows, the location of the emission at
this velocity is significantly different between the two
observations.

\subsection{Methanol Masers}
\label{sec:meth}


\begin{table}
\begin{tabular}{|c|c|c|c|c|c|c|c|c|c|}

  \hline

{NO.} & V$_{LSR}$ & Flux &  $\alpha$ & error& $\delta$ & error     \\
    & km s$^{-1}$  & Jy/b   & sec      &  sec & $''$       & $''$      \\\hline
1 & -5.51 &  1.82$\pm$0.01 &   26.0563 &  0.0002 &  32.680 &  0.002 \\
2 & -6.12 & 20.31$\pm$0.03 &   26.0516 &  0.0002 &  32.705 &  0.002 \\
3 & -6.19 &  4.47$\pm$0.03 &   26.0505 &  0.0002 &  32.760 &  0.002 \\
4 & -6.53 &  4.50$\pm$0.02 &   26.0517 &  0.0002 &  32.652 &  0.002 \\
5 & -6.48 &  3.80$\pm$0.02 &   26.0564 &  0.0002 &  32.746 &  0.002 \\
6 & -6.54 & 10.43$\pm$0.02 &   26.0570 &  0.0002 &  32.694 &  0.002 \\
7 & -6.49 &  2.15$\pm$0.02 &   26.0579 &  0.0002 &  32.639 &  0.002 \\
8 & -7.04 &  1.29$\pm$0.01 &   26.0666 &  0.0002 &  32.669 &  0.002 \\
9 & -7.17 & 10.11$\pm$0.01 &   26.0721 &  0.0002 &  32.711 &  0.002 \\
10& -7.19 &  1.73$\pm$0.01 &   26.0706 &  0.0002 &  32.765 &  0.002 \\
11& -7.09 &  0.51$\pm$0.01 &   26.0748 &  0.0002 &  32.609 &  0.003 \\
12& -7.72 &  4.24$\pm$0.03 &   26.0677 &  0.0002 &  32.794 &  0.002 \\
13& -7.75 &  3.02$\pm$0.03 &   26.0635 &  0.0002 &  32.700 &  0.002 \\
14& -7.71 &  1.25$\pm$0.03 &   26.0586 &  0.0002 &  32.613 &  0.003 \\
15& -7.71 &  0.83$\pm$0.03 &   26.0534 &  0.0003 &  32.533 &  0.003 \\
16& -7.81 &  1.04$\pm$0.03 &   26.0696 &  0.0002 &  32.685 &  0.003 \\
17& -8.40 &  2.11$\pm$0.01 &   26.0760 &  0.0002 &  32.711 &  0.002 \\
 \hline

\end{tabular}
\caption{The parameters of the components detected in CH$_3$OH masers
 towards IRAS~$20126+4104$. The leading terms  of the positions are
 $\alpha$(J2000)= 20$^h$ 14$^m$ and $\delta$(J2000)= 41$^{\circ}$ $13'$.}
\end{table}

\begin{figure}
  \centering
  \includegraphics[angle=0,width=6cm]{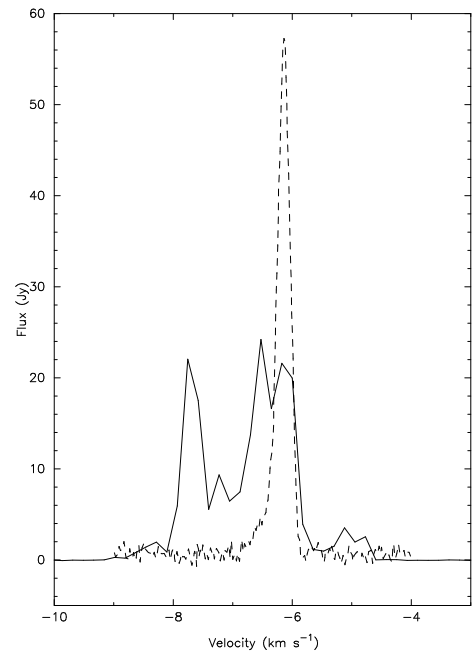}
\caption{CH$_3$OH masers spectra towards IRAS~$20126+4104$ from
MERLIN (solid line) compared with spectra of Onsala-Effelsberg
baseline (dashed line) from Minier et al.
(2001).}\label{fig:Meth_spec}
\end{figure}

Ten methanol maser features were detected, all very close to the
northern group of OH masers. Table 6 presents the parameters of
the components and Figure~\ref{fig:Meth_spec} shows the methanol
masers spectrum
compared with that detected by Minier et al. (2001). The methanol
spectra has changed significantly since the observations of
Minier et al. (2001), as has also been shown recently by Galt (2004).
 Also the MERLIN
observations show the maser emission originates much closer to the
central source than the EVN map of Minier et al.(2001).

\section{Discussion}

The exact location of the central driving source in IRAS~$20126+4104$
is uncertain. It is unclear whether the double radio continuum source
detected by Hofner et al. (1999) represents the outflow on either side
of the source, or whether the brighter, south eastern component is
coincident with the driving source as suggested by a model for the
\htwoo\ maser emission developed by MCR. The observations of the dust
continuum emission do not provide any definitive support for either
interpretation as the absolute positional uncertainty of the 1.3mm and
3mm observations, $\sim0.9''$, is too large to distinguish between
these two possibilities. However, it is clear from our MERLIN
observations that all three types of maser originate in the inner
circumstellar region, within $\sim0.5''$ (850 AU) of the central
source.

\subsection{The Disk}

As discussed above, the OH masers are confined in two clumps,
while the \methanol\ masers are closely associated with the
northern of these clumps. The axis of the elongated distribution
of maser spots is perpendicular to the axis of the jet from the
central source (C99). The same orientation has been inferred for a
circumstellar disk around the central source from observations of
CH$_3$CN, H$^{13}$CO$^+$ (C97, C99) and NH$_3$ (Zhang et al.
1998).  These molecular line observations all identify a velocity
gradient along the proposed disk, and the \methanol\ and OH maser
velocities appear generally consistent with these velocity
gradients. The northern OH masers, all right-hand circularly
polarised, and have velocities ranging from -14~km s$^{-1}$ to
-12~km s$^{-1}$, while the southern right-hand polarised
components have velocities of 2.1~km s$^{-1}$. The remaining two
maser spots, both associated with the southern clump, have
velocities of $\sim-4$~km s$^{-1}$, although it should be noted
that one of these spots is clearly a component of a Zeeman pair
with one of the right-hand spots. 

\begin{figure}
  \centering
  \includegraphics[angle=0,width=6cm]{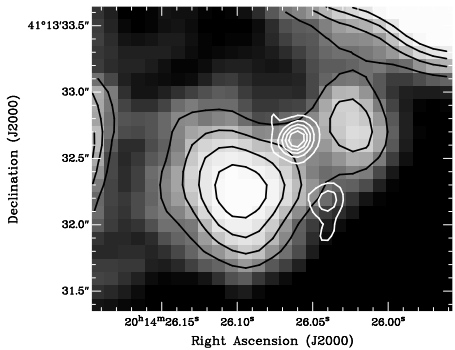}
\caption{The integrated OH maser emission (bold contours)
superimposed
  on an image of the K (2.2-$\mu$m) image of the source (also seen in
  thin contours), from Sridharan et al. (2004). The OH masers clearly
  arise from the
  material giving rise to the dark extinction lane between the two lobes
  of the nebula. The extinction lane is due to the material in the
  circumstellar disk around the central source. The positional
  uncertainty in the IR image is estimated to 0.1$''$.}\label{fig:ir}
\end{figure}

The suggestion that the OH masers arise from material in a
circumstellar disk around the source is also supported by
comparing the OH masers with a recent high resolution infrared
image of this source. Figure~\ref{fig:ir} shows the integrated OH
emission overlaid on a 2.2$\mu$m K band image of the central
region of the source (Sridharan et al 2004). The image shows a
small bipolar K nebula with the emission lobes separated by a dark
extinction lane. This structure is identical to the kinds of
bipolar infrared nebulae seen around young low mass stars where
the K emission is due to scattering off the walls of the cavity
cleared by the outflow from the central star and the extinction is
tracing the location of a circumstellar disk around the source.
Clearly the OH masers are associated with the material responsible
for the extinction, the material presumably in a circumstellar
disk.

\begin{figure}
  \centering
  \includegraphics[angle=270,width=9cm]{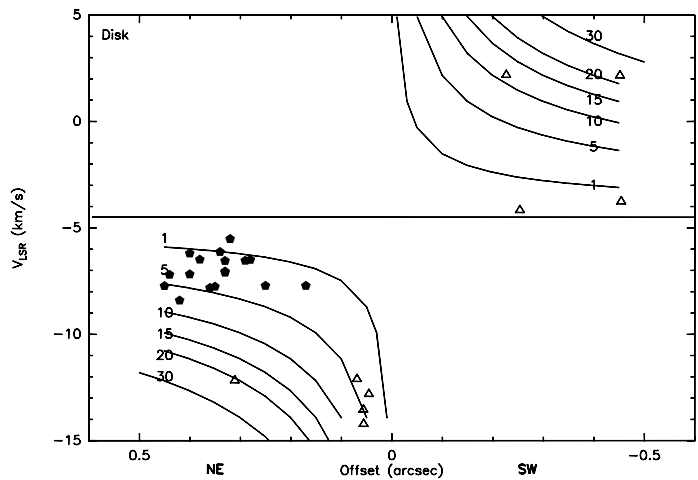}
\caption{The OH (open symbols) and methanol (filled symbols) maser
  velocities against the corresponding offset along the direction of
  the jet axis. The offset is computed with respect to the position of
  the centre of the OH masers cluster.
  The curves represent the velocities for Keplerian motion around a
  central mass of 1, 5, 10, 15 20 and 30 \Msun\ as indicated. The
  systemic velocity is taken to be 4.5 km s$^{-1}$ (Cesaroni,
  priv.comm.).}\label{fig:kepler}
\end{figure}

A position velocity diagram along the disk axis is shown in
Figure~\ref{fig:kepler}. As indicated on the diagram, most of the
OH and \methanol\ maser velocities are all consistent with a
Keplerian velocity gradient about a central source of mass $\sim$5
or $\stackrel{<}{\sim}$10~\Msun, but some of them suggest a mass of
$\sim$20~\Msun. The first value is similar to the $\sim$7~\Msun
derived by Cesaroni et al. (Cesaroni priv. comm.), while the last
value is similar to the 24~\Msun\ derived by Cesaroni et al. (1999)
and the 20~\Msun\ derived by Zhang et al. (1998). It therefore appears
that the observations suggest that these masers are arising on or, in
the surface of a circumstellar disk around this massive young star.

The presence of a OH Zeeman pair provides a measurement of the
magnetic field in this disk. The 6.3km s$^{-1}$ splitting of this
pair implies a magnetic field strength of 10.7 mG (Elitzur 1996),
pointing away from the observer. This is relatively high for
typical magnetic field strengths towards OH masers in star forming
regions (e.g. Garcia-Barreto et al. 1988; Hutawarakorn, \& Cohen
2003 and references therein).

The spatial association of the OH and 6.7-GHz methanol masers
confirms the close association of the OH and class II methanol
masers proposed by Caswell (1996) and modelled by Cragg et al.
(2002). The flux ratio of these masers, S(6668)/S(1665)= 20.3/2.7 =
7.5, places IRAS~$20126+4104$ in OH-favoured sources.
Note that although the OH and \methanol\ masers are closely associated,
there is a clear difference in position, amounting to 0.1 arcsec (170 AU),
suggesting that the OH and \methanol\ masers are not co-propagating.

Nevertheless the models of Cragg et al. (2002) can be used to
place some constraints on the physical conditions in the regions
where the masers are observed. The presence of 1665 MHz maser
emission from OH, but the absence of 1667 MHz maser emission
suggests that the material has a relatively high gas temperatures,
$T_{\rm gas} > 30$K and relatively high H$_2$ gas densities,
$n>10^6$\cc, unless the dust temperature is
$\stackrel{>}{\sim}300$K, which at the location of the OH masers
is possible, but unlikely. This also agrees with the Gray et el
(1991) model which predicted the absence of 1667 masers at high
densities. To obtain maser emission in both OH and methanol, the
Cragg et al. model requires a ratio of methanol to OH column
densities in the range 0.1 to 1000, reducing to a range from 1 to
100 if the lines are both saturated, with a ratio of 10 being
typical. Since maser emission from methanol requires methanol
abundances of greater than $10^{-6}$, this suggests an OH
abundance of $\stackrel{>}{\sim}10^{-7}$. For a methanol favoured
source such as $20126+4104$, the models suggest the gas has a
density in the range $10^{6.5}$\cc\ to $10^8$\cc\ and a gas
temperature $>125$K with dust at a temperature between 130 K and
230 K, a range of parameters consistent with the absence of 1667
MHz OH masers. The gas temperature is also consistent with the
150-260 K inferred by C97 from observations of CH$_3$CN.

The high abundance of methanol suggested by the presence of the
methanol masers could result from the liberation of methanol from
grains mantles as the dust is heated or shocked. This would also
increase the OH abundance as a result of injection of \htwoo\
followed by protonation and dissociative recombination.  The `hot
core' chemistry which results when grain mantles are evaporated
has recently be modelled by Nomura, \& Millar (2004).  This model
also shows that the gas phase abundance of CH$_3$CN is also
significantly enhanced when grain mantles are evaporated, a result
also consistent with the observations IRAS~20126+4104 (C97; C99).

\subsection{The Jet/outflow}

The water masers arise from a region spatially distinct from the
OH and \methanol\ masers. As suggested by MCR, the location of the
\htwoo\ emission appears to associate these masers with the
outflow from the source.  Figure~\ref{fig:h2o_comp} plots the location
(and velocities) of our water maser observations compared with those
detected by MCR. This figure shows that the spatial and velocity
distribution of the maser spots has changed significantly between
the two sets of observations. Indeed, it is difficult to identify any
common maser spots.

MCR proposed a detailed model for the water masers
distribution assuming that the masers lie on the surface of a
conical bipolar jet, at the interaction zone between the ionised
jet and the surrounding neutral medium, and moving with constant
velocity away from the vertex of the cone (assumed to coincide
with the embedded YSO). This model provides a prediction of the
maser velocities at any given position, and so we can compare the
measured velocities of the maser spots detected with MERLIN with
the model predictions. Doing this, we find that the model predicts
some velocities close to those observed by a value of 1.17 km
s$^{-1}$, but some predicted model velocities are higher or lower
than observed ones by value of 7 to 10 km s$^{-1}$. For example,
for the maser spot at $20^h 14^m 26^s.027$ and $41^{\circ} 13'
32''.59$, the model predicts a velocity of -4.13 km s$^{-1}$
whereas the measured velocity is $-2.27$ km s$^{-1}$. Note that
before making this comparison, the systematic velocity of $-3.5$
km s$^{-1}$ , used by MCR, was added to the velocity obtained from
Eq.(1) of the MCR model.

If we assume that the \htwoo\ masers trace a moving shock front
then they have travelled a distance of $\sim 0.1''$ over the 1587
days between our MERLIN observations and the VLBA observations of
MCR. This corresponds to a velocity of $\sim$ 190 km s$^{-1}$.
This velocity is consistent with the range of velocities inferred
for the SiO by C99. It may also they are a new masers excited by
the travelled shock. If we consider the uncertainty of the
absolute position between for our spots, 10 mas, and MCR ones, 30
mas, some of the similar velocities spots of the two observations
could be overlaped.

IRAS~20126+4104 has two different outflow directions:  a NW-SE
flow on an angular scale of $\sim$10$''$, seen in HCO$^{+}$ and
H$_{2}$ emission (C97) with an SiO jet (C99), and a large-scale CO
outflow on a 2$'$ angular scale, that is almost N-S, at position
angle 171 degrees (Shepherd et al. 2000).  Shepherd et al.
attribute the difference in position angles to precession of the
jet.  Our MERLIN measurements of the magnetic field direction show
position angles of -5$^{\circ}\pm2^{\circ}$ and
-11$^{\circ}\pm4^{\circ}$ that agree with the position angle of
the large-scale CO outflow.  (We assume that we are seeing
$\sigma$-components.)  The MERLIN measurements also agree in
position angle with the magnetic field determination by
Vall\'{e}e, \& Bastien (2000), who found a position angle of
+4$^{\circ}\pm29^{\circ}$ based on 760-$\mu$m continuum
measurements at 14$''$ resolution.

\subsection{Comparison with Other Sources}

The observations presented here show that IRAS~20126+4104 joins a
small group of luminous ($\sim10^4$\Lsun) young sources where the
OH masers originate from within $\sim1000$ AU of the central star
and often have a spatial (and in some cases kinematic) morphology
consistent the masers being located in, or on the surface of, a
circumstellar disk around the source.  The other such objects are
G35.2-0.74N (Hutawarakorn, \& Cohen 1999), W75N (Hutawarakorn et
al. 2002) and IRS1, 9 and 11 in NGC7538 (Hutawarakorn, \& Cohen
2003). The OH masers towards W3(OH) have also recently been
interpreted as originating from a circumstellar disk (Wright,
Gray, \& Diamond 2004). Although the OH masers towards
$21026+4104$ are among the weaker in this group, the 11mG magnetic
field strength measured is the highest among the sample, and is
also higher than any of the 100 field values for OH masers studied
by Fish et al. (2002).

Caswell (1996, 1998)  has suggested that OH 1665 MHz maser flux
compared to \methanol\ flux may be an indicator of the
evolutionary stage of a source, increasing as a source evolves.
This would suggest that IRAS~$20126+4104$ is amongst the younger
of these sources, which could be consistent with the presence of
1667 MHz maser emission towards the sources with stronger 1665 MHz
emission suggesting the OH emission arises from lower density
material than towards IRAS~$20126+4104$ (Cragg et al.2002).

\section{Conclusions}

  We have used MERLIN to study the immediate vicinity of
  IRAS~$20126+4104$ at high angular resolution and have shown that
  the 1665 MHz OH, \htwoo\ and \methanol\ masers towards this source
  all originate within $\sim0.5''$ (850 AU) of the central source.
  The OH masers have an elongated distribution, tracing part of a
  disk of material around the source which is orthogonal to the axis
  of the jet from the soure.  We could identify one Zeeman pair of
  OH masers which indicates a magnetic field of strength $\sim$11 mG
  in this disk. The velocity structure of the OH masers is
  consistent with Keplerian motion around a central source of
  $\stackrel{<}{\sim}$10~\Msun. The methanol masers are
  intermingled with the north-western part of the OH maser
  distribution and are at velocities intermediate between the
  north-western OH masers and those to the south-east.  Our
  observations confirm the close association of OH and methanol masers.
  We suggest that the high methanol (and OH) column densities
  necessary for the maser emission may result from the release of
  mantles from the dust grains in the surface layers of the
  circumstellar disk as the disk material has been heated by the
  central young star or as the stellar wind has shocked the disk
  material.

  The \htwoo\ masers have significantly varied since they were last
  observed at high angular resolution. We detect only one of the three
  clusters previously seen. Although the \htwoo\ masers
  detected are close to the location central source, as was seen in
  the previous observation, the maser spots have a considerably
  different spatial and kinematic structure to those previously
  measured. In particular the detailed model proposed by MCR for the
  \htwoo\ masers arising at the survey an outflow cavity fails to
  acount for the velocities and locations of the current spots.

  These observations show that, at least for this source, the three
  common types of maser associated with young high mass stars probe
  different components of the circumstellar environment allowing a
  coherent view of the circumstellar regions to be constructed. The
  OH masers provide a measurement of the magnetic field in the
  circumstellar disk within $\sim500$AU of a young high mass star.

\vspace{1cm}
{\bf ACKNOWLEDGMENTS}

We thank Peter Hofner for providing us with the 3.6cm map, and
Riccardo Cesaroni, Malcolm Gray and Anita Richards for useful communication.
MERLIN is a national facility operated by the University of Manchester
at Jodrell Bank Observatory on behalf of PPARC.

\end{document}